\documentclass[12pt]{article}  
\usepackage{amsmath}
\usepackage{amsfonts}

\newcommand{\pa}{\partial}
\newcommand{\tr}{\hbox{tr}}
\newcommand{\comment}[1]{}

\newcommand{\ncp}{b}

\newcommand{\field}[1]{\mathbb{#1}}
\newcommand{\BC}{{\field C}}
\newcommand{\CM}{{\cal M}}
\newcommand{\CO}{{\cal O}}
\newcommand{\CG}{{\cal G}}
\newcommand{\CL}{{\cal L}}

\newcommand{\plops}{plops}

\newcommand{\uiaddress}
{{\small\it Department of Physics, University of Illinois, Urbana, IL 61801}}
\newcommand{\email}[1]{\thanks{e-mail: \tt#1}}
\newcommand{\preprint}
{\begin{flushright}\begin{small}
ILL-(TH)-00-11\\ hep-th/0102158\\ 
\end{small}\end{flushright}       
}

\begin{document}

\begin{titlepage}
        \title{
        \preprint\vspace{1.5cm}
		Observations on non-commutative field theories in coordinate space}
		\author{
        	David Berenstein,\email{berenste@pobox.hep.uiuc.edu}\;
 and    
		Robert G. Leigh\email{rgleigh@uiuc.edu}\\ 
	\uiaddress
        \\
		}
\maketitle

%
%
\begin{abstract}
We discuss non-commutative field theories in coordinate space. To do so
we introduce pseudo-localized operators that represent interesting
position dependent (gauge invariant) observables. The formalism may
be applied to arbitrary field theories, with or without
supersymmetry.

The formalism has a number of intuitive advantages. First it makes clear
the appearance of new degrees of freedom in the infrared. Second, it
allows for a study of correlation functions of (composite) operators.
Thus we calculate the two point function in position space of the
insertion of certain composite operators. We demonstrate that, even at
tree level, many of the by now familiar properties of non-commutative
field theories are manifest and have simple interpretations. The form of
correlation functions are such that certain singularities may be
interpreted in terms of dimensional reduction along the non-commutative
directions: this comes about because these are theories of fundamental dipoles.
\end{abstract}

\end{titlepage}

\section{Introduction}

A great deal of effort has recently gone into the study of field
theories on non-commutative spaces,
partly due to the realization that non-commutative geometries are an
important aspect of D-brane physics \cite{CDS,SW}. There are many
interesting features, including a relation between UV and IR physics
\cite{MVRS,VRS,MST}.
A basic problem in any field theory is that of identifying the physical
observables. In the case of gauge theories on non-commutative spaces,
this is of particular interest and there are proposals for a class of
such operators built out of Wilson lines.\cite{IIKK,RU, DR, HIG} As well, there
have been many discussions of the renormalization properties of these
theories (see for example Refs. \cite{BS,VRS,MVRS,CR,Zan}).

In this paper, we will have several goals. A primary motivation for us in
thinking about non-commutative field theories has been to come to some
understanding of the
UV/IR correspondence. In particular, non-commutative field theories
apparently force us to give up on our Wilsonian thinking that is so
familiar in standard field theories. What should replace our Wilsonian
intuition? If Wilsonian
reasoning fails, in what sense can we demonstrate the renormalizability or
even the existence of a field theory?

To begin such explorations, we build field theory observables directly
in position space. In ordinary regulated field theories, local operators
should be thought of as smeared over a region whose size is
determined by the regulator. Beginning with this idea, we find that it
is possible to carry it over to the non-commutative case. An advantage
of this formalism is that it is straightforwardly extended to theories
with gauge symmetry and supersymmetry. There are aspects to this problem
which have been discussed previously; in particular, the non-commutative
coordinates are covariantized with respect to the symmetry
group\cite{MSSW,BLP}. Thus in a non-commutative field theory, the notion
of ``local operator'' is replaced by integrals over non-commutative
space with covariantized smearing distributions. We will refer to these
operators as {\it pseudo-local}.

Having constructed pseudo-local operators, one can consider the
coordinate space correlation functions. These theories are characterized
by the correlation functions of operators inserted at various positions,
and since position is not an invariant concept in non-commutative field
theories, we need to decide on a choice of correlation functions. It is natural
to take these to
correspond to the insertion of pseudo-local operators. We note that this is
the most useful setup in order to study the gravity--field theory
correspondence for these theories.\cite{M,Witten, GKP}

As we stated, a motivation to understand these correlation functions comes from
an attempt to define these theories at the quantum level. 
In a Wilsonian setup, a field
theory is renormalizable if there is a UV fixed point of the
renormalization group. Otherwise we just have an effective field theory
which is valid below some scale, and all possible operators that are
consistent with the symmetries of the theory are generated by
integrating out the heavy modes. Given the non-commutativity scale as
the origin for the UV/IR correspondence, one might wonder if this
approach makes any sense at all beyond the non-commutativity scale. One
may need new degrees of freedom in the ultraviolet. In the Wilsonian
approach, these new degrees of freedom can be integrated out and their
effects will be apparent in the effective action.

The existence of
this UV fixed point may be studied by looking at the universality of the
singularities in the correlation functions of composite operators. As
these theories are non-commutative, one can expect that as field
theories they should correspond to some new universality class in the
UV. Our results show that these theories are truly non-local in the
variables used to study them. We find singular behavior in correlation
functions at large distances in the non-commutative directions and short
distances in commutative directions, even in the free field limit. These
are long-range correlations in the non-commutative directions. Given
these correlations, one has a very hard time imagining  how to perform a
`block spin' renormalization procedure. The geometric picture of the RG
flow as exploring shorter distances on a lattice breaks down in the four
dimensional sense. Thus, non-commutative field theories are not
Wilsonian in a standard sense. The passage from coordinate to momentum space
is non-trivial in the sense that
large distance does not necessarily correspond to small momentum.

The paper is organized as follows. In Section \ref{sec:gobs} we describe
how to construct gauge invariant observables which are position
dependent. We then describe how to extend this formalism to the
construction of chiral gauge invariant observables. Algebraic details may be
found in an appendix.
In Section \ref{sec:mt} we describe how to construct multi-trace
operators which are position dependent, and we  comment on how to
interpret effective actions with insertions of these operators. The net
result is that they can not be considered to be `local' actions, and to
change them into this form it is necessary to introduce auxiliary
fields. These auxiliary fields are determined by a commutative geometry
associated to the non-commutative geometry, and thus may be
interpreted as auxiliary closed string states.\cite{SW,VRS}

Next, in Section \ref{sec:cplops}, we discuss quantum aspects of the
insertion of single trace operators, how certain divergences may be
removed and how the single trace composite operators inevitably mix with
multitrace operators (if we attempt to interpret renormalization in a
Wilsonian setting).

In Section \ref{sec:coplop} we study the correlation functions of two simple
operators inserted at different places. We show that the correlation functions can have 
ultraviolet divergences at long distances on non-commutative coordinates, and short 
distances in the commuting coordinates. These singularities are reminiscent of
 dimensionally reduced theories with a continuum spectrum of massive states (such as
in a non-compact compactification). 

\section {Gauge invariant observables: an algebraic approach}\label{sec:gobs}

In a commutative field theory, we are interested in gauge invariant observables;
in particular, we usually think of inserting operators at specific values of the
coordinates and thus build local operators. For non-commutative field theories,
there is no invariant meaning of placing an operator at $x$, as the coordinates
do not commute. 

Let us begin with a few comments on commutative field theories. Even in
a (regulated) commutative field theory, one cannot place operators
closer than the scale at which the regularization is performed. It is
more natural instead to take operators which do not have
$\delta$-function support at the point, but are rather smeared over a
non-locality scale set by the regulator, and they are thus insensitive
to the ultraviolet beyond the regularization scale. A natural operator
defined near a point $y$ in a regulated theory, $\tilde \CO(y)$, is thus
understood as a convolution of an operator $\CO(x)$ with a distribution
$f$ whose support is of a scale set by the regulator. Hence we write
 \begin{equation}
 \tilde \CO(y) = \int dx\ \CO(x) f(x-y)
 \end{equation}
with $f$ a normalized (smooth)  distribution. Removing the regulator is
equivalent to localizing $f$ towards a distribution with
$\delta$-function support.

Now, this construction may be carried over directly to non-commutative
field theories where $x$ is non-commutative: the allowable functions of
$x$ are the smooth distributions, and there is an integration operation.
We introduce a distribution $f$ of compact support (vanishing
sufficiently fast at infinity), defined through a power-series expansion
and an ordering prescription. The resulting operators will be referred
to as pseudo-localized, or simply by the acronym {\it \plops}. 

For gauge field theories, the above procedure in non-commutative
variables does not work directly as $x$ is not gauge invariant, but may
be modified very simply. We simply replace $x_j$ by 
$\hat x^j=x^j+ \ncp^{ij} A_j(x)$,
which transforms covariantly under gauge transformations (see the appendix 
for a more complete discussion\footnote{This replacement
has been also discussed elsewhere; see Refs. \cite{MSSW,BLP}.}). Notice that
the $x$ are unbounded operators and $A$ is compact, so the spectrum of
$\hat x$ is essentially the spectrum of $x$ perturbed by a small amount.
Thus, any integral which converges for $x$ will have the same
asymptotics as the integral written with $\hat x$ replacing $x$.

Now since $\hat x$ transforms covariantly under gauge transformations, we can
replace
\begin{equation}
f(x-y)\to f(\hat x - y) 
\end{equation}
Here $y$ is a parameter and is taken as a commutative variable. It is a
parameter in the distribution formal power series, and may be moved
around by translation $x\to x-a$. Since $\hat x$ transforms in the
adjoint of the gauge group, the above procedure is well defined if we
also trace over the gauge group after inserting $f(\hat x- y)$. All
operators which we can integrate are then in the adjoint of the gauge
group.  The definition of a pseudo-local operator is thus
\begin{equation}
\tilde \CO_f(y) = \int dx\ \tr\ \CO(x)f(\hat x - y)
\end{equation}
for any $\CO(x)$ which transforms in the adjoint representation, and any choice
of distribution $f$. (The $\star$-product should be understood throughout.)
We will also occasionally refer to these operators as single-trace.

Now, in order to define these operators, we have had to make a choice
of distribution $f$.  The construction has the virtue that in
the limit $\ncp\to 0$, we recover the gauge invariant commutative \plops,
 and changing $f$ will amount to a change of regularization in the
resulting commutative field theory. 
In general, there is an issue of overcompleteness. In practice, 
one would make some suitable choice of distributions $f$, perhaps chosen
to be as localized as possible (such as a coherent state or $\delta$-function
distribution). 

We have discussed here the construction of gauge-invariant operators in
position space. It is important to compare this construction to 
the other examples of gauge invariant observables that are known, such
as those involving open Wilson loops
\cite{IIKK,RU,DR,HIG}. In fact, the Wilson line construction is a special case of
the above formalism, where we take\footnote{We
thank A. Hashimoto for raising this question and helping us in the computation.
See also Ref. \cite{BLP}.}
\begin{equation}
f_k(y) = \exp(ik\cdot y).
\end{equation}
Notice that $f_k$ is maximally non-local and gives rise to the Fourier transform
necessary for a formulation in momentum space. For many applications however,
it is more appropriate to work directly in coordinate space. We will see
several such applications later in this paper. In the following subsection,
we note that the formalism may be carried over in a simple fashion to chiral
superspace.

\subsection {Supersymmetry and Chiral observables}
\label{sec:susy}

In this section, we demonstrate that the formalism presented above may
be supersymmetrized. A chiral state is annihilated by half of the
supersymmetries, say $\bar Q$. In ordinary superspace, recall that the
function $x^{\mu -}= x^\mu+i \theta \sigma^\mu \bar\theta$ is chiral, as
the variation of $x^\mu$ under supersymmetry is compensated by the
second term. Thus any power series in $x^-,  \theta$ is annihilated by
$\bar D$. This may be extended to pseudo-local operators by replacing
$x$ by $x^-$; if the operator $\CO$ is chiral, then the integral
\begin{equation}
\tilde \CO(y,\theta,\bar\theta) = 
\int dx\ \CO(x,\theta,\bar\theta)f(y- x^-)
\end{equation}
will be chiral. This is because $\bar D$ acts as a derivation, and thus
satisfies the Leibnitz rule.

Now, we would like to produce chiral gauge invariant observables in
non-commutative field theories. For the non-commutative case we had to
modify $x$ to make a gauge covariant operator $\hat x$. A naive
replacement of $x^- \to x^- + \ncp\cdot A$ gives us a gauge covariant
operator, but we lose the chirality of the integral. However, we may
build a superspace function $\hat x^-$ which is both gauge covariant and
chiral by a simple additional modification. We will present the result 
in the constrained superfield formalism.\cite{West}

We will study this by first introducing conjugate variables so that $\bar D$ may be
written in terms of a commutator. This is quite natural in the case of 
non-commutative geometry. The conjugate spinors will be referred to as 
$\eta_\alpha$, $\bar\eta_{\dot\alpha}$, and satisfy 
\begin{eqnarray}
\{\eta_\alpha, \theta^\beta\} = i \delta_\alpha^\beta\\
\{\bar\eta_{\dot\alpha}, \bar\theta^{\dot\beta}\} = i \delta_{\dot\alpha}^{\dot\beta}
\end{eqnarray}
They (anti)-commute with all the other (extended) 
superspace variables.

The superspace chiral derivatives become 
\begin{equation}
D_{\dot\alpha}\Phi = -[(\bar\eta_{\dot\alpha}
+i(\theta \sigma^{\mu})_{\dot\alpha}
p_\mu),\Phi]_\pm,\ \ \ 
D_\alpha = -[(\eta_{\alpha}
+i(\sigma^{\mu}\bar\theta)_{\alpha}
p_\mu),\Phi]_\pm\label{eq:D}
\end{equation}
Let us write the spinors on the right hand side of eqs.
(\ref{eq:D}) as $\bar\eta'_{\dot\alpha}$, $\eta'_\alpha$. 
Chiral operators are defined as those which commute with $\bar
\eta'_{\dot\alpha}$. The $\bar\eta'$ do not commute
with $x$, but they do commute with $x^{\mu -}= x^\mu+i \theta \sigma^\mu
\bar\theta$ and with $\theta$.

Covariantization procedes by introducing gauge superfields
\begin{equation}
\hat {\eta}'_{\alpha} = \eta'_{\alpha} +\Gamma_{\alpha},\ \ \ \ \ \
\hat {\bar\eta}'_{\dot\alpha} = \bar\eta'_{\dot\alpha} +\bar\Gamma_{\dot\alpha}
\ \ \ \ \ \
\hat p_\mu=p_\mu+\Gamma_\mu
\end{equation}
These are constrained by requiring that the algebra is unmodified
\begin{equation}
\{ \hat{\bar\eta}'_{\dot\alpha}, \hat {\bar\eta}'_{\dot\beta}\} = 
\{ {\hat\eta}'_{\alpha}, {\hat\eta}'_{\beta}\}=0,\ \ \ \ \ \ 
\{\hat\eta_\alpha, \hat{\bar\eta}'_{\dot\alpha}\}=-2\sigma^\mu_{\alpha\dot\alpha}\hat p_\mu
\label{eq:chcon}\end{equation}
We will solve these constraints directly. In the non-commutative case, we
must also covariantize
\begin{equation}
\hat x^{\mu -}=x^{\mu -}+\ncp^{\mu\nu}\tilde\Gamma_\nu
\end{equation}
We will require that this is chiral
\begin{equation}
[\hat{\bar\eta}'_{\dot\alpha},\hat x^{\mu -}]=0
\label{eq:chtil}\end{equation}
and gauge covariant.

We will work at linearized order, although it is straightforward to
generalize. The constraints (\ref{eq:chcon}) may be solved by
\begin{equation}
\Gamma_\alpha=[\eta'_\alpha,\Omega]\ \ \ \ \ \ 
\Gamma_{\dot\alpha}=[\bar\eta'_{\dot\alpha},\bar\Omega]
\end{equation}
We then find
\begin{equation}
\Gamma_\mu=-i\pa_\mu\bar\Omega+\frac{1}{4}\bar\sigma_\mu^{\dot\alpha\alpha}
[\bar\eta'_{\dot\alpha},[\eta'_\alpha,\Omega-\bar\Omega]]
\end{equation}
These are standard results and are unmodified here.  

The chirality constraint (\ref{eq:chtil}) has solution
\begin{equation}
\tilde\Gamma_\nu=-i\pa_\nu\bar\Omega+\Phi_\nu
\end{equation}
where $\Phi_\mu$ is covariantly chiral. 
The superfield $\Phi_\mu$ may be chosen such that the lowest component of
$\tilde\Gamma_\mu$ is $A_\mu$. 

These results may be presented in the Wess-Zumino gauge; another possibility
is to use a complexified gauge transformation to set $\bar\Omega=0$, the 
chiral gauge.
In this gauge, the chirality constraint is unmodified, and we find that
$\hat x^{\mu -}=x^{\mu -}+b^{\mu\nu}a_{\nu}(x^-,\theta)$, explicitly chiral.
This is covariant under a restricted class of chiral gauge transformations.
In more general gauges, a gauge transformation is required to bring $\hat x^-$
back to this form.

These results should allow calculations to be done in supersymmetric
non-commutative field theories. The single trace operators presented
above are expected to correspond to supergravity fields in the large
$N$ limit, by the AdS/CFT \cite{Witten}. Also, the formalism is important
in that the renormalization properties, and existence, of interacting
supersymmetric theories are expected to be in better shape.

\section{Multi-Trace Operators and ``Effective'' Actions}\label{sec:mt}

The operators that we have discussed above are by construction single trace. 
We can
algebraically construct multi-trace pseudo-local operators by taking products of
the single trace \plops. These are given by
\begin{equation}
\tilde \CO_{MT, f_1, f_2, \dots f_n} (y) = 
\tilde \CO_{1f_1}(y) \tilde \CO_{2f_2} (y)\dots \tilde \CO_{n f_n}(y) \label{eq:simple}
\end{equation}
If written in terms of integrals over the non-commutative space, this
operator looks non-local as we get multiple integrals, one for each
trace.\footnote{ One can construct more general multitrace pseudolocal
operators by further smearing the relative insertion of the operators,
but the spirit of the construction of operators is the same.} This
discussion is at this point classical, but we will return to quantum
issues later.

It is interesting to contemplate the fate of a Wilsonian formalism in
this coordinate space language. To this end, we might suppose that we
have a notion of an effective action at a given momentum scale.
As we evolve this scale, the generic form of the action will involve
an infinite set of \plops, {\it including multi-trace operators}. This
is inevitable: we expect this in ordinary gauge theories and
therefore in non-commutative gauge theories as well. In fact, these multi-trace
operators are the source of the UV-IR correspondence and the apparent
loss of a Wilsonian action. 

The point is as follows. Since the effective action is non-local 
(in the strongest sense we are defining locality by requiring the number 
of non-commutative integrals to be equal to one), it is
natural to attempt to make it look local at the expense of introducing
new degrees of freedom. In the case of double-trace operators, this is
particularly simple: the degrees of freedom are ``wormhole parameters.''
A simple pseudo-local double-trace operator takes the form
\begin{equation}
\int dy \tilde \CO_{12}(y) = \int dy \int dx \CO_1 f_1(\hat x- y) 
\int dx\CO_2 f_2(\hat x - y)
\end{equation}
We may introduce auxiliary fields as Lagrange multipliers in the path
integral such that the action can be written
\begin{equation}
\int dy b_1(y) b_2(y) - i\int dy b_1(y) \tilde \CO_1(y)- 
i\int b_2(y) \tilde \CO_2(y) 
\end{equation}
and a functional integral over the $b(y)$ is understood. Notice that the
$b$ are functions of commutative variables $y$, and one can think of
$\int dy b_1(y) \hat f(x-y)= \tilde f_b(\hat x)$ as a new distribution
with which we are convoluting the operator $\CO_1$.
As the $b$ see a commutative geometry and as they couple to single trace
operators in the field theory, one can think of them as closed string
auxiliary fields by combining the ideas of \cite{Witten,SW}. This result
matches nicely with the infrared divergences encountered in \cite{VRS,MVRS}.

Thus the action for double-trace operators is of the form
\begin{equation}
\int dy b_1(y) b_2(y) + \int dx \CO_1(x) \tilde f_{1,b_1}(\hat x) 
+\int dx \CO_1(x)\tilde f_{2,b_2}(\hat x)
\end{equation}
which looks local from the non-commutative field theory perspective as
having one integral over non-commutative space, and hence can be interpreted as
 a 'local' action. Notice that in this manner, operators
which we would have called non-local can be made to look local in the
sense of having only one integral over the non-commutative plane at the 
expense of introducing new degrees of freedom.

For triple trace operators and beyond one can add extra polynomial terms 
in the $b(y)$. Upon integrating out the auxiliary fields
we will be left with a power series in multitrace operators.

\section{Understanding composite \plops}\label{sec:cplops}

The analysis we have done so far is classical: the operators
$\tilde\CO(y)$ need to be defined at the quantum level. In the
definition of $\tilde\CO(y)$ we made the substitution $x\to \hat x$,
which involves the connection. As the operators are all inserted at the
same point (since in a single trace operator we only integrate over one
variable) we expect to find divergences that would normally be
associated to contact terms. The structure of these divergences is also
apparent in the work of \cite{HIG} where they were understood in
terms of the diagrammatic expansion of the Wilson loop in a power series
in the connection. Thus to define these operators quantum-mechanically
we need to regularize the operators in some scheme. (In supersymmetric
field theories these difficulties might be avoided if we use operators
which are protected by supersymmetry. The formalism of Section 
\ref{sec:susy} would be helpful in such computations.)

There are really two separate issues here. We can distinguish the
singularities of the operator ${\cal O}$ (perhaps fixed via a normal
ordering procedure) from that of the singularities between two single-trace
operators.

In this section we will try to understand the structure of \plops\ which
have self-contractions. This is necessary if we want to understand the
renormalizability of a bare action and the structure of counterterms
required to render a theory finite at a given order in perturbation
theory.
There is a point to be mentioned here which is important. The connection
enters in the definition of the operators, and if we want to calculate 
Feynman diagrams when we contract the connection, the propagators
carry a factor of $g^2$. Thus, if we are renormalizing by taking power 
series in the couplings, only a finite number of contractions are
needed at each order in $g$. For our calculation we will take the leading 
tree level contributions, and hence the connection will not play any role at all.

We begin by considering the operator $\phi^2(y)$. In a commutative theory,
we might define this by point-splitting, subtracting the short-distance pole.
More generally, we can write
\begin{equation}
\phi(y)^2=\int dz\ K(z,y) \phi(z)\phi(y)
\end{equation}
for some suitable kernel $K$. For example, we could choose the kernel such that
the short-distance singularity is integrable. In the non-commutative case, we 
interpret this as a definition of
$(\tr\phi)^2$.

As for $\tr\phi^2$, we would, at least classically, have written
\begin{eqnarray}
\phi^2(y)&=&\int dx\ \phi^2(x)f(x-y)\\
&=& \int dx\ \int dk_1\ dk_2\  e^{ik_1\cdot x} e^{ik_2\cdot x}
 \phi(k_1)\phi(k_2)f(x-y)
\end{eqnarray}
We propose that quantum mechanically, we write the same expressions.
With some manipulations, we have
\begin{equation}
\phi^2(y)=\int dx\ \int dk_1\ dk_2\ e^{i(k_1+k_2)\cdot x}e^{ik_1\wedge k_2}
f(x-y)\phi(k_1)\phi(k_2)
\end{equation}
As momenta are commutative parameters, we can subtract the pole in 
$\phi(k_1)\phi(k_2)$ as in commutative field theory. Thus we write
\begin{equation}
\phi^2(y):=\int dx\ \int dk_1\ dk_2\ e^{i(k_1+k_2)\cdot x}e^{ik_1\wedge k_2}
f(x-y):\phi(k_1)\phi(k_2):
\end{equation}
It is important to note that composite operators thus defined already contain
Moyal phases.

Other composite operators may be normal-ordered in
a similar fashion and include in general both planar and non-planar
subtractions. For example, consider the composite pseudo-local 
operator $\phi^4(y)$ (within a $\phi^4$ theory). We start with
\begin{eqnarray}
\phi^4(y) &=& \int dk_1\ dk_2\ dk_3\ dk_4\ \tr
\left[ \phi(k_1)\phi(k_2)\phi(k_3)\phi(k_4)\right]\\
&\cdot&
e^{i(k_1+k_2+k_3+k_4)\cdot y}\ e^{ik_1\wedge k_2+i(k_1+k_2)\wedge k_3
+i(k_1+k_2+k_3)\wedge k_4}
\end{eqnarray}
For simplicity we will take the distribution $f$ to be a $\delta$-function.

After a little work, we find
\begin{eqnarray}
\tr(\phi^4)(y) &=& :\tr(\phi^4(y)):+\hbox{planar contractions}\nonumber\\
&&+\int d^d k  \left(\frac{1}{k\circ k+1/\Lambda^2}\right)^{1-d/2}
\tr(\phi(k))\tr(\phi(y+\ncp\cdot k)) e^{ik \cdot y}
\end{eqnarray}
The planar terms are proportional to $:\tr\phi^2(y):$. In the non-planar
term that we have written explicitly, we have dropped finite numerical
constants for simplicity and taken the mass $m\to 0$.

The point that we wish to emphasize here is that double-trace operators
necessarily appear in the definition of composite operators. If we attempted
to interpret these equations as the starting point for a renormalization
group approach to composite operators, then we conclude that single trace
\plops\ mix with multi-trace \plops. This same behaviour occurs if we
attempt to construct a Wilsonian action--we are unable to write a local, 
single-trace action.

\section{Correlation functions of \plops}\label{sec:coplop}

In this section, we will consider tree level correlation functions of
pseudo-local operators. We will find that much of what is known about
non-commutative field theories is already visible at tree level--one does
not need to look at loop effects to uncover the relevant physics. We will
consider correlation functions $
\langle\phi^n(0) {\bar{\phi}}^n(y)\rangle$.\footnote{Again, we will
simplify the computations by taking $f(x-y)=\delta(x-y)$. Also, we now
consider complex scalar fields.}

Let us first consider the correlation function with $n=1$,
$\langle\tilde \phi(0) \tilde{\bar\phi}(y)\rangle$.
A simple computation shows that this is planar, and thus is the
same in both commutative and non-commutative theories.

Now consider the composite operator $\phi^2(0)$. As discussed in the last
section, we have
\begin{equation}
\phi^2(0)= \int d k_1\ dk_2\ e^{i k_1\wedge k_2} \phi(k_1)\phi(k_2)
\end{equation}
Now we want to calculate the correlation between this operator and
$\bar\phi^2(y)$ which is similarly defined:
\begin{equation}
\bar\phi^2(y)= \int dk'_1\ dk'_2\ e^{i k'_1\wedge k'_2}\ \bar\phi(k'_1)
\bar\phi(k'_2)\ e^{i (k'_1+k'_2)\cdot y}
\end{equation}
Clearly, the correlation function has two distinct contractions. One of
these is planar while for the non-planar contraction we find:
\begin{equation}
\int dk_1\ dk_2\ \frac{
e^{2ik_1\wedge k_2}e^{-i(k_1+k_2)\cdot y}}{(k_1^2+m^2)(k_2^2+m^2)}
\end{equation}
Regularizing,\footnote{As is standard, the short distance regularization
is done by modifying propagators $\frac1{k^2+m^2} \to \int_0^\infty
d\alpha e^{-\alpha(k^2+m^2)-1/(\Lambda^2\alpha)} $.} we can rewrite this
as
\begin{equation}
\int d\alpha\ d\beta \int dk_1\ dk_2\ 
e^{2ik_1\wedge k_2}e^{-i(k_1+k_2)\cdot y}e^{-\alpha(k_1^2+m^2)}e^{-\beta(k_2^2+m^2)}
e^{-\frac 1{\Lambda^2\alpha}-\frac1{\Lambda^2\beta}}
\end{equation}
Now, we do the integrals over $k_1, k_2$, and we separate the distance
into $a = (a_{NC}, r)$, to distinguish separation along the
non-commutative directions and along the commutative directions. We will
assume that there are precisely two non-commutative directions; the
generalization is straightforward. As the integral is Gaussian, it is
easy to perform and (modulo numerical factors) gives
\begin{equation}
\int d\alpha d\beta
\frac1{\alpha\beta+\ncp^2}\frac1{(\alpha\beta)^{(d-2)/2}} 
e^{-\alpha m^2-\beta m^2-\frac{\alpha+\beta}{\alpha\beta+\ncp^2} \frac{a_{NC}^2}{4} -
\left( \frac{1}{\alpha}+\frac{1}{\beta}\right)\left(\frac{r^2}{4}+\frac{1}{\Lambda^2}\right)}
\end{equation}

As $\ncp\to 0$ while holding $\Lambda$ fixed, this goes smoothly to the
commutative field theory limit. For finite $\ncp$, we can examine the
behavior of the integral using a saddle-point approximation, and
consider various limits. From the form of the integral it is clear that
$\Lambda$ and $1/r$ behave the same way: thus large momenta and small
commutative distance are the same. Distance in the non-commutative
directions behaves quite differently however. In particular, at large
$b$, $a_{NC}$ no longer acts like a UV cutoff, but instead
$a_{NC}^2/4\ncp^2$ behaves as a mass.

The nature of the saddle point depends on  whether $\alpha, \beta$ are
large or small with respect to the non-commutative parameter. If
$\alpha,\beta$ are large, then $\ncp$ is not important for the
evaluation of the saddle point, and the result reproduces the commutative field
theory limit. This is true if $\frac{r^2}{4}+\frac{1}{\Lambda^2}$ is
large in units of the non-commutative distance.

If $\alpha, \beta$ are very small, then $\ncp$ is important, and the
relevant saddle point is obtained by the Taylor expansion of
$\frac{\alpha+\beta}{\alpha\beta+\ncp^2} \frac{a_{NC}^2}{4}$ around
$\alpha, \beta \sim 0$, and we get instead a contribution to the mass
which is proportional to $a_{NC}^2$.
The asymptotic form of the expression is proportional to
\begin{equation}
\langle\phi^2(0) \bar\phi^2(a,r) \rangle
\sim
\frac{e^{-2r\sqrt{m^2+a^2/4\ncp^2}}}{ b^2 r^{d-4}}
\end{equation}
Thus for $r$ small (even when $a_{NC}$ is very large) 
the correlation function has  singular behaviour no matter what
the value of $a_{NC}$ is, although it is  suppressed as we let
$a_{NC}$ get larger.

This
singularity is like the one corresponding to a $d-2$ dimensional field
theory, as if the non-commutative directions are compactified, and with
an effective mass given by $m^2_{eff} =m^2+a^2/4\ncp^{2}$. The fact that
this singularity is present signals that there are long-distance
correlations along the non-commutative directions. In an ordinary field
theory  long-distance correlations are attributed to massless modes
propagating in these directions, thus it is not surprising that when we
do loop diagrams we get 'infrared' singularities that would normally be
associated to the effects of propagation of such modes.

The effective two-dimensional mass associated to a mode with momentum $
k $ along the non-commutative directions is $m^2+k^2$. Thus the behavior
of the correlation function is associated to a mode such that $\ncp
k\sim a$. Indeed, for the rigid dipole picture \cite{BS,SJ} of the non-commutative
modes, the effective distance between the ends of the dipoles at such
momentum between points zero and $a$ can be approximately zero. This is
how we can interpret the above singularity. The UV degrees of freedom
are large, they grow in size with the momentum. Thus localized
(regularized) operators become larger in size as we send the cutoff
scale to infinity. 

It should be noted that the theory is more complicated than a simple
dimensional reduction however. There are still the planar graphs which behave like
ordinary particle graphs in $d$ dimensions. The correct statement is that
the degrees of freedom $\phi(k)$ are dipoles.

This is perhaps the clearest way in which we can see that the theory
fails to be Wilsonian. As we increase the momentum, we are not
necessarily exploring shorter distances, so the connection between the
regulator and the coarse graining of the geometry fails. The structure
of singularities fails to be universal in the standard sense. We  find
UV divergences at long distances, where we expect only IR physics.

Notice that the above calculation is independent of whether the theory
is supersymmetric or not. Even with a lot of supersymmetry, these strong
correlations will not disappear; instead the behavior of the correlation
functions will be protected by supersymmetry. If we take the cutoff to
be related to a very small momentum scale with respect to the
non-commutativity scale, the effects of the non-commutative parameter
are unimportant, and the theory behaves like ordinary $d$-dimensional
field theory. That is, the $\alpha, \beta\to 0$ singularity is cut off by
the regulator scale, and as $\alpha, \beta$ are large compared to
$\ncp$, $\ncp$ is not important. In this sense we can recover ordinary
field theory results in the infrared if we take the momentum cutoff
scale to zero. This is in accordance with the supergravity solutions
corresponding to $N=4, d=4$ SYM on a non-commutative plane \cite{HI,MR}.

Clearly, the interesting behavior is for large $\Lambda$, when one is
trying to understand the ultraviolet structure of the theory. The main
result here is that the singularity structure makes part of the
theory behave like a dimensionally reduced field theory. 

If one is to do a Wilsonian analysis at all, it might be best 
to consider it only along the directions that are Lorentz
invariant. The non-commutative directions should be treated as being
compactified on a non-compact space, in a spirit similar to the
Randall-Sundrum scenarios \cite{RS}, as the mass parameter $a$ is a
continuous one. In this vein notice that for four dimensions, we get a
singularity that behaves like $\log (r)$ at short distances. If we
integrate over $r$ to make a term in the effective action the measure
will render these terms finite, so one can still call the theory 
renormalizable,  but not in a Wilsonian interpretation.

\section{Conclusion}

In this paper we have given a procedure to build gauge invariant
operators on a non-commutative field theory where the operators are position
dependent, that is, they have a
local character. We have also extended this construction to build
chiral pseudo-local  gauge invariant observables in supersymmetric 
non-commutative field theories.

The construction of states gives an overcomplete basis of operators, and
also allows us to build multi-trace pseudo-local observables. In other
treatments of these multi-trace  operators higher $\star_n$ products are
introduced\cite{DasT,L,LM,MW}. We feel that the
presentation given here is intuitively more appealing, although it is
not clear that one can write an  effective action economically. This
depends on the choice of basis for the functions which one uses in the
convolutions.

We have studied some correlation functions of these operators at the
free field theory level. Already these exhibit the UV/IR correspondence.
We have found that there are `non-planar' contributions to correlation
functions of operators which are separated by large distances in the
non-commutative directions and short distance in the commutative
directions which give rise to UV singularities. These behave as if the
theory is (partially) dimensionally reduced. As such, the theories fail to behave
like standard Wilsonian field theories,  the structure of UV
singularities in correlation functions of operators do not conform to
the geometrical picture of block-spin  renormalization. The
singularities in principle can give rise to infinities and
renormalization of multi-trace operators when we integrate over the
position of these operators, but we have found no such behavior in four
dimensions. For higher dimensional field theories one can get UV
singularities this way, but the theories are already
non-renormalizable and one needs to introduce new physics in the
ultraviolet of the theory in any case.

Given the complicated structure of the correlation functions, one might
still ask if there is some other  universal singular behavior in the
correlation functions which might give us some operational definition of
non-commutative field theories and their renormalizability. This new
structure might account for a new way to understand operator product
expansions and rescue in some sense the Wilsonian approach to
renormalization, which has proved so useful in other contexts, and which
at this moment seems unable to cope with the problem of non-commutative
field theories. These issues are currently under investigation and a
more detailed account will appear elsewhere \cite{BLprog}.

\bigskip
\noindent {\bf Acknowledgments:} We wish to thank E. Fradkin, A. Hashimoto, M. van Raamsdonk,
S. Shenker and L. Susskind for valuable discussions.
Work supported in part by U.S.
Department of Energy, grant DE-FG02-91ER40677. 
\appendix
\section{Appendix: Algebraic notes}

We are considering non-commutative field theories
with spacelike non-commutativity in four dimensions.
 The non-commutativity is given by 
\begin{equation}
[x^i, x^j] = i \ncp^{ij}
\end{equation}
with $\ncp^{ij}$ antisymmetric and real.

It is well-known that functions
are then multiplied using the Moyal product
\begin{equation}
f(x)\star g(x) = \exp(-i\ncp^{ij}\partial_i\partial_j')
f(x) g(x')|_{x'=x}.
\end{equation}
One should be careful to consider what are allowable functions on
the
non-commutative space. The $x$ are coordinates of the non-commutative 
space, and
they generate the ring of functions. However, the $x$ are not in the ring of
allowable functions on the space, as they correspond to unbounded
non-normalizable operators. We take the allowable functions to be the ring of
formal power series in the $x$ which are convergent and decay sufficiently fast
at infinity, that is, a Schwarz space for the algebra generated by the $x$. For
this to be a $\BC^*$ algebra, we need to complete it. The functions can be
represented by operators in a Hilbert space ${\cal H}$, and they are all compact.
The completion is done with the following norm
\begin{equation}
\langle f,g\rangle = \int d^2 x f^\star(x) g(x) = \tr_{\cal H}(\hat f^\dagger\hat g) 
\end{equation}
where $\hat f, \hat g$ are the representative operators of the functions
$f,g$. We require $|f|<\infty$ with the above norm, so the operators in the 
ring of functions are square integrable.
In particular, the ring of operators contains all finite matrices.


%

We denote by $G$ a finite Lie group with Lie algebra $\CG$, which is finite
dimensional. On a commutative space $\CM$ we can consider tensoring the algebra
of smooth functions on $\CM$ with $\CG$. We call these the infinitesimal gauge
transformations. As the functions commute, the commutator of two infinitesimal
gauge transformations closes on the Lie algebra of infinitesimal gauge
transformations.

On a non-commutative space one can try to mimic this construction, but one runs
into problems. Let us consider this possibility. The infinitesimal gauge
transformations are given by sums of the form
\begin{equation}
\CL(x) = \sum_a f_a(x) \CL_a = f_a(x) \CL_a
\end{equation}
with $\CL_a$ the generators of the Lie algebra $\CG$. We think of $\CL(x)$ as an
operator acting on some Hilbert space and the Lie bracket 
is the commutator of
two of these. We must interpret $\CL$ in terms of the enveloping algebra of $\CG$
in order to make sense of the product.

The commutator of two of these generators is given by
\begin{eqnarray}
[\CL,\CL'] &=& f_a(x) \CL_a f'_b(x) \CL_b-f'_b(x) \CL_b f_a(x) \CL_a\\
&=&\frac 12( f_a(x)\star f'_b(x)+ f'_b(x) \star f_a(x))[\CL_a,\CL_b]\\
&&
+\frac12( f_a(x)\star f'_b(x) - f'_b(x)\star f_a(x))(\{\CL_a,\CL_b\})
\end{eqnarray}
now, $[\CL_a, \CL_b] = f_{abc} \CL_c$, so the first term in the above expression
is in the Lie algebra, but the second one can only be in the Lie algebra if
$\{\CL_a,\CL_b\} \in \CG$. Usually this term belongs to the enveloping algebra of
$\CG$, but not to the Lie algebra and thus there is a technical problem in
defining non-commutative gauge theories.

This problem may be resolved by using the enveloping algebra of $\CG$ and
requiring a finiteness condition on the representation. For simple $\CG$ it acts
on the field  as the ring of $n\times n$ matrices for some $n$, and the gauge
group is $U(n)$.\footnote{Other gauge groups may also be obtained, but they
require a more complicated structure \cite{BSST}}

Fields transform as irreducible representations of the Lie algebra either by
matrix multiplication on the left or the right. Thus the only allowable
representations have to be either left modules of the ring of $n\times n$
matrices tensored with the algebra of allowable functions, or they are right
modules or bimodules of this algebra. That is, the  fields transform in the
fundamental, antifundamental or adjoint of the gauge group.

If we forego the  $\CG$  simple condition, then one can get product groups
$\prod_i U(n_i)$. For each $U(n_i)$ factor a field can transform as a left module
or a right module, but it cannot be a left module under two of the $U(n_i)$, as
the $x$ do not commute. Thus matter can only transform in adjoints under one
group, and as left-right modules under two different gauge groups. This is
exactly the type of spectrum that one can get for D-branes at singularities, so
any non-commutative field theory is described by some quiver diagram (not
necessarily associated to an orbifold). For each node in the quiver diagram we
have an associated projector $P_i$ which commutes with the Lie algebra of gauge
transformations, and projects onto the direct summand associated to the node.

The gauge transformations thus act on fields as
\begin{equation}
\delta\phi = \CL_L\star \phi - \phi\star \CL_R
\end{equation}
where $L,R$ denotes the transformation under the gauge group operating
on the left or right of the field $\phi$ belonging to the 
$(n_L,\bar n_R)$ 
representation of the gauge group, and the matrix multiplication is implicit.

The covariant derivative will be given by a commutator 
\begin{equation}
D_j\phi = \partial_j \phi - [i A_j,\phi] 
\end{equation}
with $A_j$ a Lie algebra valued allowable function. This is a compact way of writing
\begin{equation}
D_j\phi = \partial_j \phi - iA_{jL} \phi + i\phi A_{jR} 
\end{equation}

In terms of the projectors for the nodes in the quiver diagram associated to 
the field theory
we have
\begin{equation}
P_L \phi = \phi P_R = \phi
\end{equation}
for a field $\phi$ which corresponds to an arrow from the node associated to $P_L$ to
the node associated to $P_R$
and also we have
\begin{equation}
A_L = P_L A = A P_L 
\end{equation}
for the connections and we can do similarly for gauge transformations.

Consider the case
where the only non-trivial commutation relation is $[x^2, x^3] = i \ncp$. 
The derivative operators in the $2,3$-directions are equivalent to
commutators
\begin{eqnarray}
D_3 \phi &=& \frac{1}{-i\ncp}[\phi,x_2 + \ncp A_3]\\
D_2 \phi &=& \frac{1}{i\ncp}[\phi,x_3 - \ncp A_2]
\end{eqnarray}
The operators $\hat x^\mu=x^\mu +\ncp^{\mu\nu} A_\nu$ are gauge
covariant (they transform in the adjoint representation), which one may show
explicitly by direct computation. The non-homogeneous term in the transformation
of $A$ is compensated because $x$ does not commute with the gauge
transformations.
The commutator of these gives us
\begin{equation}
[\hat x_2, \hat x_3] = i\ncp - i \ncp^2 F_{23}
\end{equation}
with $F$ the curvature of the gauge group in the $2,3$ directions, which is
 an allowable
function. It can be shown that
\begin{equation}
[D_2,D_3] (\phi) = i\left( F_{23} \phi- \phi F_{23}\right)
\end{equation}
The variables $\hat x_j$ are the tools that we need to build gauge invariant observables.

\providecommand{\href}[2]{#2}\begingroup\raggedright\endgroup

\end{document}